\documentclass[12pt]{article}
\input tcilatex
\QQQ{Language}{
American English
}

\begin{document}

\author{A.I.Volokitin$^{1,2}$ and B.N.J.Persson$^1$ \\
%EndAName
\\
%EndAName
$^1$Institut f\"ur Festk\"orperforschung, Forschungszentrum \\
J\"ulich, D-52425, Germany\\
$^2$Samara State Technical University, 443100 Samara,\\
Russia}
\title{Non-contact friction between nanostructures}
\maketitle

\begin{abstract}
We calculate the van der Waals friction between two semi-infinite 
solids in normal relative motion and find a drastic difference in 
comparison with the parallel relative motion. 
The case of the good conductors is investigated in details both within 
the local optic approximation, and using a non-local optic 
dielectric approach. 
We show that the friction may increase  by many order of
magnitude when the surfaces are covered by adsorbates, or can support
low-frequency surface plasmons. In this case
the friction is determined by resonant photon tunneling between adsorbate
vibrational modes, or surface plasmon modes.
The theory is compared to atomic force
microscope experimental data.
\end{abstract}

\section{ Introduction}
A great deal of attention has been devoted to non-contact friction between
nanostructures, including, for example, the frictional drag force between
two-dimensional quantum wells \cite{Gramila1,Gramila2,Sivan} , and the
friction force between an atomic force microscope tip and a substrate \cite
{Dorofeev,Gostmann,Stipe,Mamin,Hoffmann}.

In non-contact friction the bodies are separated by a potential barrier
thick enough to prevent electrons or other particles with a finite rest mass
from tunneling across it, but allowing interaction via the long-range
electromagnetic field, which is always present in the gap between bodies.
The presence of inhomogeneous tip-sample electric fields is difficult to
avoid, even under the best experimental conditions \cite{Stipe}. For
example, even if both the tip and the sample were metallic single crystals,
the tip would still have corners present and more than one crystallographic
plane exposed. The presence of atomic steps, adsorbates, and other defects
will also contribute to the inhomogeneous electric field. The electric field
can be easily changed by applying a voltage between the tip and the
sample.

The electromagnetic field can also be created by the fluctuating current density,
 due to
thermal and quantum fluctuations inside the solids. This fluctuating
electromagnetic field is always present close to the surface of any body,
and consist partly of traveling waves and partly of evanescent waves which
decay exponentially with the distance away from the surface of the body. The
fluctuating electromagnetic field originating from the fluctuating current
density inside the bodies gives
rise to the well-known long-range attractive van der Waals interaction
between two bodies \cite{Lifshitz}. If the bodies are in relative motion, the
same fluctuating electromagnetic field will give rise to a friction which is
frequently named as the van der Waals friction.
 Van der Waals friction can be
considered as mediated by photon exchange between the bodies: One body emit
a photon, and the other absorbs it, thus transferring momentum between the
bodies, resulting in a friction force. At large distances between the
bodies, the main contribution to friction comes from photon exchange,
corresponding to the propagating electromagnetic waves. However this
contribution is very small because the photons corresponding to propagating
waves carry very small momentum, no larger than  $ k_BT/c\hbar $. The photons,
corresponding to the evanescent electromagnetic waves, carry the momentum $%
q\sim d^{-1}$. Thus for distances $d$ between two bodies smaller
characteristic distance $d_T=\hbar c/k_BT$, which depends on temperature (at
room temperature $d_T\sim 10^5$\AA ), the main contribution to friction
comes from the evanescent electromagnetic field. In analogy with electron
tunneling, this mechanism of momentum transfer can be considered as
associated with the photon tunneling.

Although the dissipation of energy connected with the non-contact friction
always is of electromagnetic  origin, the detailed mechanism is not
totally clear, since there are several different mechanisms of energy
dissipation connected with the electromagnetic interaction between bodies.
First, the electromagnetic field from one body will penetrate into the other
body, and induce an electric current. In this case friction is due to Ohmic
losses inside the bodies. The fluctuating electromagnetic  
 field can also excite the vibrations of the adsorbates or other 
surface localized modes, e.g. surface plasmons and polaritons.
In this case friction is due to energy relaxation of the surface modes.
 Another contribution to friction from the
electromagnetic field is associated with the time-dependent stress acting
on the surface of the bodies. This stress can excite acoustic waves, or
induce time-dependent deformations which may result in a temperature
gradient. It can also induce motion of defects either in the bulk, or on the
surface of the bodies. The contribution to friction due to non-adiabatic heat
flow, or motion of defects, is usually denoted as internal friction.

It is very worthwhile to get a better understanding of different mechanisms
of non-contact friction because of it
practical importance for
ultrasensitive force detection experiments. This is because the ability to
detect small forces is inextricably linked to friction via the
fluctuation-dissipation theorem. For example, the detection of single spins
by magnetic resonance force microscopy, which has been proposed for
three-dimensional atomic imaging \cite{Sidles} and quantum computation \cite
{Berman}, will require force fluctuations to be reduced to unprecedented
levels. In addition, the search for quantum gravitation effects at short
length scale \cite{Arkani} and future measurements of the dynamical Casimir
forces \cite{Mohideen} may eventually be limited by non-contact friction
effects.

Recently Gotsmann and
Fuchs  \cite{Gostmann} observed long-range non-contact friction between an aluminum
tip and
a gold (111) surface. The friction force $F$ acting on the tip is
proportional to the velocity $v$, $F=\Gamma v$. For motion of the tip
normal to the surface
the  friction coefficient $\Gamma (d)=b\cdot d^{-3}$,
where $d$ is the tip-sample spacing and $b=(8.0_{-4.5}^{+5.5})\times
10^{-35}\mathrm{N\,s\,m}^2$  \cite{Gostmann}. Later Stipe \textit{et.al.}\cite{Stipe}
observed non-contact friction effect between a gold surface and a gold-coated
cantilever as a function of tip-sample spacing $d$, the temperature $T$, and the
bias voltage $V$. For vibration of the tip parallel to the surface they
found $\Gamma (d)=\alpha (T)(V^2+V_0^2)/d^n$, where $n=1.3\pm 0.2,$ and $%
V_0\sim 0.2\,\mathrm{V.}$ At 295\textrm{K, }for the spacing $d=$ 100\textrm{%
\AA\ }they found $\Gamma =1.5\times 10^{-13}\mathrm{\,kgs}^{-1},$ which is $%
\sim $500 times smaller that reported in Ref. \cite{Gostmann} at the same
distance using a parallel cantilever configuration.

In a recent Letter, Dorofeev \textit{et.al.} \cite{Dorofeev} claim that a
the non-contact friction effect observed in \cite{Dorofeev,Gostmann} is due to
Ohmic losses mediated by the fluctuating electromagnetic field . This result
is controversial, however, since the van der Waals friction has
been shown \cite{Volokitin1,Persson and Volokitin,Volokitin2,Volokitin3} to
be many orders of magnitude smaller than the friction  observed by Dorofeev \textit{%
et.al.} Presently, the origin of the difference in magnitude and distance
dependence of the long-range non-contact friction effect observed in \cite
{Gostmann} and \cite{Stipe} is not well understood.

In order to improve the basic understanding of  non-contact friction,
 we present new results for van der Waals friction.
In \cite{Volokitin1} we developed a theory of van der Waals friction
for surfaces in parallel relative motion.
Here we generalize the theory to include also the case when the
surfaces are in
normal relative motion, and we show that there is drastic difference between
these two cases. Thus,  for normal relative motion of clean good conductor surfaces,
the friction is many orders of magnitude larger than for parallel relative motion,
but  still  smaller than observed  experimentally.
Another enhancement mechanism of the non-contact friction can be connected
with resonant photon tunneling between  states localized on the different
surfaces. Recently it was discovered that resonant photon tunneling between
surface plasmon modes give rise to extraordinary enhancement of the optical
transmission through sub-wavelength hole arrays \cite{Ebbesen}. The same
surface modes enhancement  can be expected  for van der Waals friction if the
frequency of these modes is sufficiently low to be excited by thermal radiation.
At room temperature only the modes with frequencies below
$\sim 10^{13}s^{-1}$ can be excited.
For normal metals surface plasmons have much too high frequencies; at
thermal frequencies
the dielectric function of normal metals becomes nearly
purely imaginary, which exclude surface plasmon enhancement of the van der Waals
friction for good conductors. However surface plasmons
for semiconductors are characterized by much
smaller frequencies and damping constants, and they can give an important contribution
to van der Waals friction.
Other  surface modes which can be excited
by thermal radiation are
adsorbate vibrational modes. Especially for parallel vibrations these modes
may have very low frequencies.

All information about the long-range electromagnetic interaction between two
non-contacting bodies is, in principle, contained in the reflection factors
of the electromagnetic field. At present time very little is known about the
reflection factors for large wave vectors and for extremely small
frequencies. In the calculations of the reflection factors one must 
take into account the non-local response of the electron gas on the external 
electromagnetic field. There are two correlation length which determine this 
nonlocal response. First is the skin depth, which determines the long-range 
length scale 
of  the nonlocality in the volume, and the second is the screening length. 
 The latter length scale determines the short range  correlation length for non-local 
response in the surface region.   
In our previous calculations of the Van der Waals friction \cite
{Volokitin1,Persson and Volokitin,Volokitin2,Volokitin3} we mostly
considered good conductors. In this case it was shown that the important
contribution comes from the non-local optic effects in the surface region.
However it was shown that the Van der Waals friction becomes much larger for
high resistivity material, for which the volume contribution from non-local
effects is also important. It is easy to see that within local optic
approximation the Van der Waals friction diverge when the conductivity of
materials tend to zero. This means that the local optic approximation breaks
down for high-resistivity materials. This situation is completely different
from the heat transfer between bodies via photon tunneling \cite
{Pendry,Volokitin2}, where the heat flux is maximal at conductivities
corresponding to semi-metals. In order to clarify the situation we study the
dependence of the van der Waals friction on the dielectric properties of the
materials within the non-local dielectric approach, which was proposed some
years ago for the investigation of the anomalous skin effects \cite{Fuchs
and Kliever}.

\section{Calculation of the fluctuating electromagnetic field}

 We consider two semi-infinite metals
\textbf{1} and \textbf{2} having parallel flat surfaces. We introduce a
coordinate system with $xy$ plane in the surface of body \textbf{1} , 
 and the $z
$ axis along the upward normal. The surface of body \textbf{2} is located at
$z=d$, performing small amplitude vibrations along the $z$ axes with
displacement coordinate $u_z(t)=u_0e^{-i\omega _0t}.$ Since the system is 
 translation invariant in the $\mathbf{x}=(x,y)$ plane, the electromagnetic
field can be represented by the Fourier integral
\begin{eqnarray}
\mathbf{E}(\mathbf{x},z)=\int \frac{d^2q}{(2\pi)^2}e^{i\mathbf{q} \cdot \mathbf{x}}
\mathbf{E}(\mathbf{q},z),   \\
\mathbf{B}(\mathbf{x},z)=\int \frac{d^2q}{(2\pi)^2}e^{i\mathbf{q} \cdot \mathbf{x}}
\mathbf{B}(\mathbf{q},z),
\end{eqnarray}
where $\mathbf{E}$ and $\mathbf{B}$ are the electric and magnetic induction field,
and $\mathbf{q}$ is the two-dimensional wave vector in $(x,y)$ plane. After the
Fourier transformation it is convenient to choose the coordinate axis in the 
(x,y) plane along the vectors $\mathbf{q}$ and $\mathbf{n}=[\hat{z}\times \mathbf{q}
]$. In the vacuum
gap between the bodies  the electric field $\mathbf{E}(\mathbf{q},\omega,z)$,
 and the
magnetic induction field $\mathbf{B}(\mathbf{q},\omega,z)$, can, to the linear order
in the vibrational coordinate, be written in the form
\begin{equation}
\mathbf{E}(\mathbf{q},\omega ,z)=\left( \left( \mathbf{w}_{0}e^{ipz}
+ \mathbf{v}_{0}e^{-ipz}\right)+\left(\mathbf{w}_{1}e^{ip^{+}z}+\mathbf{v}_{1}e^{-ip^{+}z}
\right)e^{-i\omega_0 t}
\right)
e^{-i\omega t}  \label{one}
\end{equation}
\begin{eqnarray}
\mathbf{B}(\mathbf{q},\omega ,z)=c\left[\frac 1{\omega} \left([\mathbf{k}^{-}\times
\mathbf{v}_{0}]e^{-ipz}+ [ \mathbf{k}^{+}\times\mathbf{w}_{0}]e^{ipz}\right)
 \right.  \nonumber   \\
\left. +\frac{1}{\omega +\omega _0}\left(
[ \mathbf{k}^{-}_+\times\mathbf{v}_{1}]e^{-ip^+z}+ [\mathbf{k}^{+}_+
\times\mathbf{w}_{1}e^{ip^+z}]\right)e^{-i\omega _0t}
\right] e^{-i\omega t}  \label{two}
\end{eqnarray}
where $\mathbf{k}^{\pm}=\mathbf{q}\pm \hat{z}p,\, 
p=((\omega/c)^2-q^2)^{1/2},\,p^+=p(\omega+\omega_0),\,\mathbf{k}^{+}_{+}
=\mathbf{k}^{+}(\omega+\omega_0),\,\mathbf{k}^{-}_{+}
=\mathbf{k}^{-}(\omega+\omega_0).$ At the surfaces of the bodies the amplitude
of the outgoing electromagnetic wave must be equal to the amplitude of the reflected 
wave plus the amplitude of the radiated wave. It is convenient to decompose the 
electromagnetic field into the $p$- and $s$ - polarized electromagnetic waves. 
For $p$-polarized electromagnetic waves the electric field is in the incident 
plane determined by the vectors $\mathbf{q}$ and $\mathbf{n}$, and for $s$- 
polarized electromagnetic waves the electric field is normal to the incident 
 Thus the boundary conditions 
for the electromagnetic 
field at $z=0$
can be written in the form
\begin{equation}
w_{0z(y)}=R_{1p(s)}(\omega )v_{0z(y)}+E^{f}_{1z(y)}(\omega )
\label{three}
\end{equation}
\begin{equation}
w_{1z(y)}=R_{1p(s)}(\omega +\omega _0)v_{1z(y)}  \label{four}
\end{equation} 
where $R_{1p(s)}(\omega )$ is the reflection factor for surface \textbf{1} 
 for $p(s)$ - polarized electromagnetic field, and where $%
E^{f}_{1z(y)}(\omega )$ are the components of the
 fluctuating electric field outside the surface \textbf{1} in the
absence of the body \textbf{2}. The boundary condition at the surface of
the body \textbf{2} must be written in the reference frame where the body \textbf{2}
is at rest. The electric field in this reference frame is determined by a Lorentz
transformation. Performing a Lorentz transformation of the electric field to linear order in
$\omega _0$ gives
\begin{equation}
\mathbf{E}^{\prime }=\mathbf{E}-\frac{i\omega _0u(t)\left[ \mathbf{\hat e}%
_z\times \mathbf{B}\right] }c  \label{five}
\end{equation}
For the $p$-polarized electromagnetic waves the second term in (\ref{five}) 
is of the order of magnitude $\omega_0 u_0 \omega/pc^2$ relative to the first one and
can be neglected for the most practical cases. However, for the $s$-polarized 
electromagnetic
waves the second term is of the order of magnitude $\omega_0 u_0p/\omega$ and can be
of the same order of magnitude as the first term. In the rest frame of body \textbf{2}
there is also mixture of $s$- and $p-$ polarized electromagnetic waves. In \cite
{Volokitin1} it was was shown that this  gives contribution of the 
order $(\omega_0u_0/c)^2$  
and thus can be neglected. After performing Lorentz transformation to linear order in 
$\omega_0$ and $u_0$ we get 
$ \mathbf{v}_{0}^{\prime}=
\mathbf{v}_{0},\,\mathbf{w}_{0}^{\prime}=\mathbf{w}_{0}$
\[
v_{1z(x)}^{\prime}=v_{1z(x)}-ipu_0v_{0z(x)};\,\,
w_{1z(x)}^{\prime}=w_{1z(x)}+ipu_0w_{0z(x)};
\] 
\[
w_{1y}^{\prime}=w_{1y}+\frac{\omega +\omega_0}{\omega}ipu_0w_{0y};\,\,
v_{1y}^{\prime}=v_{1y}-\frac{\omega +\omega_0}{\omega}ipu_0v_{0y}
\]
The boundary conditions for the electromagnetic field at  $z=d+u(t)$ in the rest
frame of body \textbf{2}
can be written in the form
\begin{equation}
v_{0z(y)}=e^{2ipd}R_{2p(s)}(\omega )w_{0z(y)}+e^{ipd}E^{f}_{2z(y)}(\omega )
\label{six}
\end{equation}
\begin{equation}
v_{1z}-ipu_0v_{0z}=e^{2ip^{+}d}R_{2p}^{+}(w_{1z}+ipu_0w_{0z})
\label{seven}
\end{equation}
\begin{equation}
v_{1y}-ipu_0\frac{(\omega +\omega _0)v_{0y}}\omega =e^{2ipd}R_{2s}^{+}
\left(w_{1y}+ipu_0\frac{(\omega +\omega _0)w_{0y}}\omega\right)
\label{eight}
\end{equation}
where $R_{2p(s)}(\omega )$ is the reflection factor for surface \textbf{2} 
  for $p(s)$ - polarized electromagnetic field, and where $
E_{2z(y)}^f(\omega )$ are the components of the
 fluctuating electric field outside the surface \textbf{1} in the
absence of the body \textbf{1}.
From (\ref{three},\ref{four}) and (\ref{six}-\ref{eight}) we get
\begin{equation}
w_{0z(y)}=\frac{R_{1p(s)}E^{f}_{2z(y)}e^{ipd}+E^{f}_{1z(y)}}{\Delta}  \label{nine}
\end{equation}
\begin{equation}
v_{0z(y)}=\frac{e^{2ipd}R_{2p(s)}E^{f}_{1z(y)}+E^{f}_{2z(y)}e^{ipd}}{\Delta}  
\label{ten}
\end{equation}
\begin{equation}
v_{1z}=ipu_0 \frac{(e^{2ipd}R^{f}_{2p}+e^{2ip^{+}d}R_{2p}^{+})E^{f}_{1z}+
(1+e^{2ip^{+}d}R_{2p}^{+}
R_1)E^{f}_{2z}e^{ipd}}
{\Delta_p \Delta_p^{+}}  \label{eleven}
\end{equation}
\begin{equation}
v_{1y}=ipu_0\frac{\omega + \omega_0}{\omega}
 \frac{(e^{2ipd}R_{2s}+e^{2ip^{+}d}R_{2s}^{+})E_{1y}^f+
(1+e^{2ip^{+}d}R_{2s}^{+}R_{1s})E_{2y}^fe^{ipd}}
{\Delta_s \Delta_s^{+}}  \label{twelve}
\end{equation}
\begin{equation}
w_{1z(y)}=R_{1p(s)}^{+}v_{1z(y)}  \label{thirteen}
\end{equation}
where $R_{p(s}^{+}=R_{p(s)}(\omega +\omega_0),\, \Delta_{p(s)}=1-e^{2ipd}R_{2p(s)}
R_{1p(s)},$ and $
\Delta_{p(s)}^{+}=\Delta_{p(s)}(\omega +\omega_0)$. Other components of the 
fluctuating electromagnetic field can be found from the transversality conditions
\begin{equation}
qw_x+pw_z=0,\,\, qv_x-pv_z=0
\end{equation}
The fundamental characteristic
of the fluctuating electromagnetic field is the correlation function, determining
the average product of components $\mathbf{E}^{f}(\mathbf{q},\omega)$. 
Accordingly to the
general theory of the fluctuating electromagnetic field  
(see for a example \cite{Volokitin2}) these correlation
function are given by 
\begin{equation}
<|E^{f}_{y}(\mathbf{q},\omega)|^2>=\frac{\hbar \omega^2}
{2c^2|p|^2} \left(n(\omega)+\frac{1}{2}\right)
[(p+p^*)(1-|R_s|^2)+(p-p^*)(R_s^*-R_s)]  \label{corfs}
\end{equation}
\begin{equation}
<|E^{f}_{z}(\mathbf{q},\omega)|^2>=\frac{\hbar q^2}
{2|p|^2} \left(n(\omega)+\frac{1}{2}\right)
[(p+p^*)(1-|R_p|^2)+(p-p^*)(R_p^*-R_p)]
\label{corfp}
\end{equation}
where $<...>$ denote statistical average over the random field, 
and  where the Bose-Einstein factor
\[
n(\omega )=\frac 1{e^{\hbar \omega /k_BT}-1}
\]
 
We note that $p$ is real for $q<\omega/c$ (propagating waves), and  purely
imaginary for  $q>\omega/c$ (evanescent waves).
Thus for $q<\omega/c$  
and $q>\omega/c$ the 
correlation functions are determined by the first and the second 
terms in Eqs. (\ref{corfs})and (\ref{corfp}), respectively. 

\section{Calculation of the friction force between two semi-infinite bodies in normal
relative motion}
 
The frictional stress $\sigma$ which act on the surfaces of the two bodies
can be obtained from $zz-$ component of the Maxwell stress tensor $\sigma_{ij}$
, evaluated at $z=0$: 
\[
\sigma _{zz} =\frac 1{4\pi }\int_0^\infty d\omega \int \frac{d^2q}{(2\pi)^2}
\Big[ <\left|
E_z(\mathbf{q},\omega ,z)\right| ^2>+<\left| B_z(\mathbf{q},\omega
,z)\right| ^2> 
\]
\[ 
  -<\left| E_x(\mathbf{q},\omega ,z)\right| ^2> -<\left| E_y(\mathbf{q},\omega ,z)\right| ^2>  
\]
\begin{equation} 
 -<\left| B_x(\mathbf{q}      
,\omega ,z)\right| ^2> - <\left| B_y(\mathbf{q}                  
,\omega ,z)\right| ^2>\Big] _{z=0}   \label{3thirteen}
\end{equation}
 To linear order in the 
vibrational coordinate $u(t)$ and the frequency $%
\omega _{0\text{ }},$ the stress acting on the surface \textbf{1} can be
written in the form
\begin{equation}
\sigma _{zz}=\sigma _{0zz}(d)+u(t)\frac \partial {\partial d}\sigma
_{0zz}(d)+i\omega _0\gamma _{\perp }u(t)  \label{3fifteen}
\end{equation}
Here the first term determines the conservative van der Waals stress and the 
second term is the adiabatic change of the conservative van der Waals stress during
vibration. The last term determines the frictional stress with friction coefficient
$\gamma_{\perp}$. For normal relative
motion (see Appendix A) we obtain the friction coefficient 
$\gamma _{\perp} =\gamma_{\perp}^{rad}+\gamma_{\perp}^{evan}$,
where the contribution to the friction coefficient from the propagating
electromagnetic waves is given by
\[
\gamma_{\perp}^{rad}=\frac{\hbar}{4\pi^2}\int_0^{\infty}d\omega
\left(-\frac{\partial n}{\partial \omega}\right)\int_0^{\omega/c}dq\,qp^2
\]
\[
\times   \big[(1-|R_{1p}R_{2p}|^2)^2+|(1-|R_{1p}|^2)R_{2p}e^{ipd}
\]
\begin{equation}
+(1-|R_{2p}|^2)R_{1p}^{*}e^{-ipd}|^2\big]\frac 1{\left|1-e^{2ipd}R_{1p}R_{2p}\right|^4}
+[p\rightarrow s],  \label{3seventeen}
\end{equation} 
and where the contribution to the friction from the evanescent electromagnetic
waves is given by
\[
\gamma_{\perp}^{evan}=\frac{\hbar}{\pi^2}\int_{\omega/c}^{\infty}d\omega
\left(-\frac{\partial n}{\partial \omega}\right)\int_{\omega/c}^{\infty}dq\,
qk^2e^{-2kd}
\]
\[
\times \big[\big(\mathrm{Im}R_{1p}+
e^{-2kd}\left|R_{1p}\right|^2\mathrm{Im}R_{2p}\big)\big(\mathrm{Im}R_{2p}+
e^{-2kd}\left|R_{2p}\right|^2\mathrm{Im}R_{1p}\big)
\]
\begin{equation}
+e^{-2kd}\big(\mathrm{Im}\big(R_{1p}R_{2p}\big)\big)^2\big]
\frac 1{\left|1-e^{-2kd}R_{1p}R_{2p}\right|^4}
+[p\rightarrow s],
\label{3eighteen}
\end{equation}
where $k=|p|$. The symbol $[p\rightarrow s]$ in Eqs. (\ref{3seventeen}) and 
(\ref{3eighteen}) denotes the term which is
obtained from the first one by replacement of the reflection factors $%
R_p(\omega ),$ for $p-$ polarized waves, by the reflection factors $%
R_s(\omega )$ for $s-$ polarized waves. The friction coefficient for 
two flat surfaces in parallel relative motion was obtained by us before,  
 \cite{Volokitin1} and 
can be written as
$\gamma _{\parallel} =\gamma_{\parallel}^{rad}+\gamma_{\parallel}^{evan}$,
where the contribution to the friction coefficient from the propagating
electromagnetic waves is given by
\[
\gamma_{\parallel}^{rad}=\frac{\hbar}{8\pi^2}\int_0^{\infty}d\omega
\left(-\frac{\partial n}{\partial \omega}\right)\int_0^{\omega/c}dq\,q^3
\]
\begin{equation}
\times   \frac{(1-|R_{1p}|^2)
(1-|R_{2p}|^2)}{\left|1-e^{2ipd}R_{1p}R_{2p}\right|^2}
+[p\rightarrow s],  \label{3nineteen}
\end{equation}
and where the contribution to the friction from the evanescent electromagnetic
waves is given by
\[
\gamma_{\parallel}^{rad}=\frac{\hbar}{2\pi^2}\int_0^{\infty}d\omega
\left(-\frac{\partial n}{\partial \omega}\right)\int_{\omega/c}^{\infty}dq\,
q^3e^{-2kd}
\]
\begin{equation}
\times \mathrm{Im}R_{1p}
\mathrm{Im}R_{2p}
\frac1{\left|1-e^{-2kd}R_{1p}R_{2p}\right|^2}
+ [p\rightarrow s].
\label{3twenty}
\end{equation}
There is a principal difference
between the friction coefficient for normal and parallel relative motion,
related to the denominator in the formulas for the friction coefficient.
The resonant condition corresponds to the case when the denominator of the integrand
in Eqs. (\ref{3seventeen}-\ref{3twenty}), which is due to multiple 
scattering of the evanescent electromagnetic
waves from the opposite surfaces, is small. For two identical surfaces and
$R_{i}<<1\le R_r$,
where $R_i$ and $R_r$ are the imaginary and real part, respectively, this
corresponds to the resonant condition $R_r^{2}\rm{exp}(-2kd)\approx 1$. At
resonance the integrand in Eqs. (\ref{3nineteen}) and (\ref{3twenty}) 
has a large factor $\sim 1/R_i^2$,
in  sharp contrast to the case of parallel relative motion, where  there is no such
enhancement factor. The resonance condition can be fullfiled even for the case
when $\rm{exp}(-2kd)<<1$ because for evanescent electromagnetic waves there is no
restriction on the magnitude of the real part or the modulus of $R$.
This open up the possibility of
resonant denominators for $R_r^2>>1$.

To estimate the friction coefficient $\Gamma$ for an atomic force microscope 
tip we can use an approximate formula \cite{Hartmann,Johansson}
\begin{equation}
\Gamma=2\pi\int_0^\infty d\rho\rho\gamma(z(\rho))
\label{approx}
\end{equation}
where it is assumed that the tip has cylinder symmetry. Here $z(\rho)$ 
denotes the tip - surface distance as a function of the distance $\rho$ 
from the tip symmetry axis, and the friction coefficient $\gamma(z(\rho)$ 
is determined by the expressions for the flat surfaces. This scheme was 
proposed in \cite{Hartmann} for the calculation of the conservative van der 
Waals interaction. The error of these scheme is not larger than 5-10\% in 
practice in an atomic force microscopy experiment, and 25\% in a worst case 
satiation \cite{Johansson}. Although this scheme was proposed for the  
conservative van der Waals interaction, we assume that the same scheme is 
also valid for the calculation of the van der Waals friction. We assume that 
the tip has a paraboloid shape given [in cylindrical coordinates ($z,\rho$)] 
by the formula:$z=d+\rho^2/2R$, where $d$ is the distance between the tip 
and the flat surface, and where $R$ is the radius of curvature of the tip. 
In the case of the power dependence
\[
\gamma(\rho) = \frac{C}{\left(d+\frac{\rho^2}{2R}\right)^n}
\]
we get
\[
\Gamma=\frac{2\pi R}{n-1}\frac{C}{d^{n-1}}=\frac{2\pi Rd}{n-1}\gamma(d)
\]      
In a more  general case one must use  numerical integration.

In the local optic approximation, where the dielectric function 
is assumed to depend only on the frequency $\omega$,  
the reflection factors  $R_{p}$ and $R_{s}$ for flat surfaces,
 covered by an adsorbate layer, are 
 given by \cite{Langreth}:
\begin{equation}
R_{p}=\frac {p   -s/\epsilon-4\pi in_aq
[s\alpha_{\parallel}/\epsilon-q \alpha_{\perp}]}
{p+s/\epsilon-4\pi in_aq[s\alpha_{\parallel}/\epsilon+q \alpha_{\perp}]}, 
\label{3twentyone}
\end{equation}
\begin{equation}
R_{s}=\frac {p   -s-4\pi in_a(\omega/c)^2\alpha_{\parallel}}
{p   +s+4\pi in_a(\omega/c)^2\alpha_{\parallel}}, \label{3twentytwo}
\end{equation}
where
\begin{equation}
s=\sqrt{\left(\frac{\omega}{c}\right)^2\epsilon - q^2},   
\end{equation}
and where $\alpha_{\parallel}$ and $\alpha_{\perp}$ are the
polarizabilities of adsorbates
in a direction parallel and normal to the surface,  respectively. Here 
$\epsilon=\epsilon(\omega)$
is the bulk  dielectric function and $n_a$ is
the concentration of adsorbates. For clean surfaces $n_a=0$, and in this
case formulas (\ref{3twentytwo,3twentythree}) reduce to the well-known Fresnel formula.

At $d<l,v_F/\omega$ and $k_F\sim 1$, where $l$ is the electron mean 
free path, and where $v_F$ and $k_F$ are the Fermi velocity and Fermi wave number, 
respectively, 
the system will be characterized by 
non-local dielectric function $\epsilon(\mathbf{q},\omega)$. 
In this paper  we use
the non-local optic dielectric
approach, proposed some years ago for the investigations of the
optical properties of a semi-infinite electron gas \cite{Fuchs and Kliever}.

Accordingly  to  \cite{Fuchs and Kliever}, the reflection factor for $p$
- polarized electromagnetic field, incident on the flat surface, is
determined by \cite{Fuchs and Kliever}
\begin{equation}
R_p=\frac{p-Z_p}{q+Z_p},  \label{rone}
\end{equation}
 where the surface impedance $Z_p$ is given by
\begin{equation}
Z_p=\frac {2i}{\pi} \int_0^\infty \frac{\mathrm{d}q_z}{Q^2}\left( \frac{q^2}{%
\epsilon _l(\omega ,Q)}+\frac{(\omega /c)^2q_z^2}{(\omega /c)^2\epsilon
_t(\omega ,Q)-Q^2}\right) ,  \label{rtwo}
\end{equation}
where $\epsilon _l$ is the finite- life- time generalization of the
longitudinal Lindhard dielectric function which accordingly to \cite{Mermin}
can be written as:
\begin{equation}
\epsilon _l(\omega ,Q)=1+\frac{(1+\mathrm{i}/\omega \tau )(\epsilon
_l^0(\omega +\mathrm{i}/\tau ,Q)-1)}{1+(\mathrm{i}/\omega \tau )(\epsilon
_l^0(\omega +\mathrm{i}/\tau ,Q)-1)/(\epsilon _l^0(0,Q)-1)},  \label{rthree}
\end{equation}
\begin{equation}
\epsilon _l^0(\omega ,Q)=1+\frac{3\omega _p^2}{Q^2v_F^2}f_l,  \label{rfour}
\end{equation}
\begin{equation}
f_l=\frac 12+\frac 1{8z}\left( [1-(z-u)^2]\ln \frac{z-u+1}{z-u-1}%
+[1-(z+u)^2]\ln \frac{z+u+1}{z+u-1}\right) ,  \label{rfive}
\end{equation}
where $Q^2=q^2+q_z^2,z=Q/2k_F,u=\omega /(Qv_F)$, $\omega _p$ is the plasma
frequency, $\tau $ is the Drude relaxation time, and where $v_F$ and $k_F$
are the Fermi velocity and wave vector, respectively. For $s-$ polarization
the reflection factor is determined by
\begin{equation}
R_s=\frac{1-Z_sp}{1+Z_sp}  \label{rsix}
\end{equation}
where
\begin{eqnarray}
Z_s &=&\frac {2i}{\pi} \int_0^\infty \frac{\mathrm{d}q_z}{(\omega /c)^2\epsilon
_t(\omega ,Q)-Q^2}  \label{rseven} \\
\epsilon _t(\omega ,Q) &=&1-\frac{\omega _p^2}{\omega (\omega +\mathrm{i}%
\gamma )}f_t,  \label{reight} \\
f_t &=&\frac 38(z^2+3u^{\prime 2}+1)-\frac 3{32z}\left( [1-(z-u^{\prime
})^2]^2\ln \frac{z-u^{\prime }+1}{z-u^{\prime }-1}\right.  \nonumber \\
&&\left. +[1-(z+u^{\prime })^2]^2\ln \frac{z+u^{\prime }+1}{z+u^{\prime }-1}%
\right)  \label{rnine}
\end{eqnarray}
with $u^{\prime }=(\omega +\mathrm{i}\tau ^{-1})/(Qv_F)$. We show in Sec.4
that the maximum of the electromagnetic friction is reached for small
electron densities, where the electron gas becomes non-degenerate (the electro 
 gas is degenerate for $k_BT<<\varepsilon_F$ and non-degenerate for $k_BT\ge 
\varepsilon_F$, where $\varepsilon_F$ is the Fermi energy). For non-degenerate  
 electron gas  we use the following classical expressions for dielectric functions
\cite{Landau}
\begin{eqnarray}
\epsilon _l^0(\omega ,Q) &=&1+\left( \frac{\omega _p}{Qv_T}\right) ^2\left[
1+F\left( \frac \omega {\sqrt{2}Qv_T}\right) \right]  \label{rten} \\
\epsilon _t(\omega ,Q) &=&1+\frac{\omega _p^2}{\omega (\omega +i\gamma )}%
F\left( \frac{\omega +i\gamma }{\sqrt{2}Qv_T}\right)  \label{releven}
\end{eqnarray}
where the function $F(x)$ is defined by the integral
\begin{equation}
F(x)=\frac x{\sqrt{\pi }}\int_{-\infty }^{+\infty }dz\frac{e^{-z^2}}{z-x-i0}
\end{equation}
and $v_T=\sqrt{k_BT/m}$, where $m$ is the electron mass.

\section{The case of the good conductors}

By a well-conducting metal we mean one whose dielectric function 
 $\epsilon=1-4\pi i\sigma/\omega$ ($\sigma$ is the conductivity)
 has an absolute value much larger than unity. For good conductors 
 at thermal frequencies $R_{pi}<<1$ and $R_{pr}\approx1$. Thus an
enhancement in friction is possible only for very small $q<<1/d$ .

It is convenient to write the  friction coefficient  
for the two flat surfaces 
in the form
\begin{equation}
\gamma=\hbar\int_0^{\infty}d\omega
\left(-\frac{\partial n}{\partial \omega}\right)\left(I_{ p}
+I_{s}\right)
\end{equation}
Taking into account that $qdq=kdk$,  
 from Eq. (\ref{3eighteen}) for normal relative motion of  clean surfaces 
within 
local optic approximation  
 we get contribution to friction from evanescent $p$-and $s$-polarized electromagnetic
waves  
\[
I_{\perp p}^{evan}=\int_0^{\infty}\frac{dk}{\pi^2}\,k^5\big[\mathrm{Re}(s/\epsilon)\big]^2
\Big[\big[(k^2+|s/\epsilon|^2)\mathrm{cosh}kd +2k\big[\mathrm{Im}
(s/\epsilon)\big]\mathrm{sinh}kd\big]^2  
\]
\begin{equation}
+ (k^2-|s/\epsilon|^2)^2\Big]\frac{1}{\left|\big((s/\epsilon)^2-k^2\big)
\mathrm{sinh}kd 
 + 2ik(s/\epsilon)\mathrm{cosh}kd\right|^4}
\label{4one}
\end{equation}
\[
I_{\perp s}^{evan}=\int_0^{\infty}\frac{dk}{\pi^2}\,k^5\big[\mathrm{Re}s\big]^2
\Big[\big[(k^2+|s|^2)\mathrm{cosh}kd +2k\mathrm{Im}
s\mathrm{sinh}kd\big]^2 
\]
\begin{equation}
+ (k^2-|s|^2)^2\Big]\frac{1}{\left|\big(s^2-k^2\big)
\mathrm{sinh}kd
 + 2iks\mathrm{cosh  }kd\right|^4}
\label{4two}
\end{equation}
As $k\rightarrow 0$, there is no singularity in $I_{\perp s}^{evan}$, 
and for $I_{\perp p}^{evan}$,
for $(c/\omega)|\epsilon|^{-3/2}<d<(c/\omega)|\epsilon|^{1/2}$, than 
taking into account that in this limit 
 $\mathrm{sinh}kd\approx kd$ and $\mathrm{cosh}kd\approx 1$, we get
\[
I_{\perp p}^{evan} = 2(\omega/c)^2\zeta^{\prime}\int_0^{\infty}
\frac{dk}{\pi^2}\frac{k^5}{\left|k^2d-2i(\omega/c)
(\zeta)\right|^4}    
\]
\begin{equation}
=\frac {\omega\zeta^{\prime}}{\pi^2 cd^3}\left(\frac{\pi}{2}+\arctan
\zeta^{\prime \prime}/\zeta^{\prime}- \frac{\zeta^{\prime \prime}/
\zeta^{\prime }}{1+(\zeta^{\prime \prime}/
\zeta^{\prime})^2}\right). \label{4three}
\end{equation} 
where the surface impedance  $\zeta=\epsilon^{-1/2}=
\zeta^{\prime}-i\zeta^{\prime \prime}$.
For $d<(c/\omega)|\epsilon|^{-1/2}$, $I_s$ becomes slowly dependent on $d$:
\begin{equation}
I_{\perp s} \approx \frac{1}{8\pi^2}(\omega/c)^4|\epsilon|^2
(1.22-\ln(2d|\epsilon|^{1/2}\omega/c)  \label{4four}
\end{equation} 
For $d>(c/\omega)|\epsilon|^{-1/2}$ we get
\begin{equation}
I_{\perp s}\approx (c/\omega)^2\zeta^{\prime 2}d^{-6}
\label{4five}
\end{equation}
For the propagating electromagnetic waves, taking into account that $qdq=-pdp$, 
we get
\begin{equation}
I_{\perp p}^{rad}=(\omega/c)^2\zeta^{\prime 2}\int_0^{\omega/c}\frac{dp}{\pi^2}p^5
\frac{1+\cos^2(pd)}{|p\sin pd +2i(\omega/c)\zeta\cos pd|^4},
\label{4six}
\end{equation}
\begin{equation}
I_{\perp s}^{rad}=(\omega/c)^2\zeta^{\prime 2}\int_0^{\omega/c}\frac{dp}{\pi^2}p^5
\frac{1+\cos^2(pd)}{|(\omega/c)\sin pd+ 2ip\zeta\cos pd|^4}           
\label{4seven}
\end{equation}
For $d<(c/\omega)|\epsilon|^{-1/2}$ the contribution to friction from propagating 
wave is negligibly small in the comparison with the contribution from the 
evanescent waves. For $d>(c/\omega)|\epsilon|^{-1/2}$
 as $p\rightarrow 0$, the integral $I_{\perp s}^{rad}$ has no singularity, and 
we  get for $I_p^{rad}$ 
\begin{equation}
I_{\perp p}^{rad}\approx\frac {\omega\zeta^{\prime}}{4\pi^2 cd^3}
\left(\frac{\pi}{2}-\arctan
\zeta^{\prime \prime}/\zeta^{\prime} +\frac{\zeta^{\prime \prime}/
\zeta^{\prime }}{1+(\zeta^{\prime \prime}/
\zeta^{\prime})^2}\right). \label{4eight}
\end{equation} 
In addition, $I^{rad}$ has singularities at the other zeroes of $\sin pl$,
 i.e., near the values $p_n=n\pi/d<\omega/c$ ($n$ is an integer). In the vicinity of $p_n$, 
putting $p=p_n+p^{\prime}$, we have $\sin pd\approx(-1)^n$ and $\cos pd\approx(-1)^n$, 
\[
I_{\perp p}^{rad}\approx2(\omega/c)^2\zeta^{\prime 2}\int\frac{dp^{\prime}}{\pi^2}
\frac{1}{|p_np^{\prime}d+2i(\omega/c)\zeta|^4}     
\]
\begin{equation}
\approx \frac{p_n^4c}{8\pi^2\omega d\zeta^{\prime}}\left(\frac{\pi}{2}
-\arctan\zeta^{\prime \prime}/\zeta^{\prime}- \frac{\zeta^{\prime \prime}/
\zeta^{\prime }}{1+(\zeta^{\prime \prime}/
\zeta^{\prime})^2}\right)
\label{4nine}
\end{equation}  
The number $m$ of such contribution is obviously equal to the integer part
 of the quantity $y=\omega d/\pi c$ ($m=[y]$), so that all $p_n$ (with 
the exception of $p=0$) make a summary contribution
\[
 \frac{\pi^2c}{8\omega d^5\zeta^{\prime}}\left(\frac{\pi}{2}
-\arctan\zeta^{\prime \prime}/\zeta^{\prime}- \frac{\zeta^{\prime \prime}/
\zeta^{\prime }}{1+(\zeta^{\prime \prime}/
\zeta^{\prime})^2}\right)\sum_1^m n^4=
\]
\[
\frac{\pi^2c}{8\omega d^5\zeta^{\prime}}\left(\frac{\pi}{2}
-\arctan\zeta^{\prime \prime}/\zeta^{\prime}- \frac{\zeta^{\prime \prime}/
\zeta^{\prime }}{1+(\zeta^{\prime \prime}/
\zeta^{\prime})^2}\right)
\]
\begin{equation}
\times \left[\frac{(m+1)^5}{5}-\frac{(m+1)^4}{2}+\frac{(m+1)^3}{3} -
\frac{m}{30}-\frac {1}{30}\right]
\label{4ten}
\end{equation}
In the integral $I_{\perp s}^{rad}$, the contribution from the vicinity 
of the point $p_n$ is
\[
2(\omega/c)^2\zeta^{\prime 2}\int_0^{\omega/c}\frac{dp^{\prime}}
{\pi^2}\frac{p_n^5}{|(\omega/c)p^{\prime}d+2ip_n\zeta|^4}=
\]
\[
=\frac{\omega n^2}{8cd^3\zeta^{\prime}}\left(
\frac{\pi}{2}
-\arctan\zeta^{\prime \prime}/\zeta^{\prime}- \frac{\zeta^{\prime \prime}/
\zeta^{\prime }}{1+(\zeta^{\prime \prime}/
\zeta^{\prime})^2}\right)
\]
and consequently
\[
I_{\perp s}^{rad}=\frac{\omega }{8cd^3\zeta^{\prime}}\left(
\frac{\pi}{2}
-\arctan\zeta^{\prime \prime}/\zeta^{\prime}- \frac{\zeta^{\prime \prime}/
\zeta^{\prime }}{1+(\zeta^{\prime \prime}/
\zeta^{\prime})^2}\right) \sum_1^m n^2=
\]
\begin{equation}
\frac{\omega }{48cd^3\zeta^{\prime}}\left(
\frac{\pi}{2}
-\arctan\zeta^{\prime \prime}/\zeta^{\prime}- \frac{\zeta^{\prime \prime}/
\zeta^{\prime }}{1+(\zeta^{\prime \prime}/
\zeta^{\prime})^2}\right)m(m+1)(2m+1)
\label{4eleven}
\end{equation}
At $m>>1$, when we can assume $m\approx \pi \omega/cd$, the $s-$ and $p-$
wave contribution are approximately equal, and for the total contribution 
 from propagating electromagnetic waves in this limit we get
\begin{equation}
I_{\perp}^{rad}=I_{\perp p}^{rad} +I_{\perp s}^{rad}\approx\frac
{11\omega^4}{240\pi^3c^4\zeta^{\prime}}
\label{4twelve}
\end{equation}      
The above formulas in this Section were obtained from the general Eqs. 
(\ref{3seventeen}-\ref{3twenty}) assuming absence of spatial dispersion  
of the dielectric function. But these formulas contain only surface 
impedance $\zeta$ that describe the ratio of the tangential components of the 
           electric and magnetic fields on the boundary of the body. Thus, the 
results in this section remain in force also in the presence of spatial 
dispersion, provided only that the surface impedance of the medium is 
small enough. We would have arrived at the same formulas also if we were 
to assume from the very beginning that the Leontovich boundary condition 
$\mathbf{E}=\zeta\mathbf{H}\times\mathbf{n}$ is satisfied on the surface 
of the metal. 
 
At not too low temperatures, the impedances of metals are given by the 
formula for normal skin effect
\begin{equation}
\zeta^{\prime}=\zeta^{\prime \prime}=(\omega/8\pi\sigma)^{1/2}
\label{4thirteen}
\end{equation}
where $\sigma$ is the conductivity. In the local optic approximation it is 
assuming that there is no dependence of $\sigma$ on $\mathbf{q}$. In the 
Wien region of frequencies it is also good approximation to neglect by 
the frequency dependence of $\sigma$. In this approximation using (\ref{4three}) 
 for $\lambda_W(k_{B}T/4\pi\hbar\sigma)^{3/2}<d<\lambda_W(4\pi\hbar\sigma/
(k_{B}T)^{1/2}$ ($\lambda_W=c\hbar/(k_{B}T $)
we get 
\begin{equation}
\gamma_{\perp p}^{evan}=\hbar\int_0^{\infty}d\omega\left(-\frac{\partial n}
{\partial \omega}\right)I_{\perp p}^{evan}\approx 0.13\frac{\hbar}{d^3\lambda_{W}}
\left(\frac{k_{B}T}{4\pi\hbar\sigma}\right)^{1/2}
\label{4fourteen}
\end{equation}
For the comparison the $p$-wave contribution for parallel relative motion 
 for $d<\lambda_c,\,
(\lambda_c=c/(4\pi\sigma k_{B}T)^{1/2})$
is given by (\cite{Volokitin1,Volokitin2}) 
\begin{equation}
\gamma_{\parallel p}^{evan}\approx 0.3\frac{\hbar}{d^4}\left(\frac{k_BT}{4\pi\hbar
\sigma}\right)^2
\label{4fifteen}
\end{equation}
It is interesting to note that for normal relative motion in contrast to the parallel 
relative motion practically for all $d>0$ the main contribution to friction comes 
from retardation effects because Eqs. (\ref{4fourteen}) in contrast to 
 Eq. (\ref{4fifteen}) contains the light velocity.

From Eq. (\ref{4four}) we get $s$-wave contribution to friction 
 for $d<\lambda_c$
\begin{equation}
\gamma_{\perp s}^{evan}\approx10^{-2}\frac{\hbar}{\lambda_c^4}(3-
5\ln(2d/\lambda_c))
\label{4sixteen}
\end{equation}
For parallel relative motion the $s$-wave contribution is in two times 
smaller.

For $d>\lambda_c$, taking into account that Eq. (\ref{4five}) is valid 
only for $\omega>c^2/4\pi\sigma d^2$, we get 
\begin{equation}
\gamma_s^{evan}\approx\frac{\pi k_BT\sigma}{d^2c^2}
\label{44one}
\end{equation}

From Eq. (\ref{4twelve}) we get distance independent contribution to 
friction from propagating electromagnetic waves for $d>\lambda_W$ 
\begin{equation}
\gamma_{perp}^{rad}\approx1.9\cdot10^{-2}\frac{\hbar}{\lambda_W^3\lambda_c}
\label{4seventeen}
\end{equation}

Figures 1-2  show the calculated contribution to the friction
coefficient $\gamma $ from evanescent electromagnetic waves 
for two semi-infinite solids, with parameters chosen
to correspond to copper ($\tau ^{-1}=2.5\cdot 10^{13}s^{-1}$, $\omega
_p=1.6\cdot 10^{16}s^{-1})$ at $T=273\,K$, for parallel (Fig.1) and normal 
(Fig.2) relative motion. Results are shown separately for
both the $s$- and $p$- wave contribution. The dashed line show the result
when the local (long-wavelength) dielectric function $\epsilon (\omega
)=\epsilon _l=\epsilon _t$ is used, where
\begin{equation}
\epsilon (\omega )=1-\frac{\omega _p^2}{\omega (\omega +\mathrm{i}\tau ^{-1})%
}
\end{equation}
In this case the integration in (\ref{rtwo}) and (\ref{rseven}) can be
performed analytically and we get Fresnel formulas.
Fig. 1 shows that the non-local optic effects become important 
for parallel relative motion for the $p-$
wave contribution for sufficiently small separations ($d<1000\,\mathrm{\AA }$%
). However, for the $s-$ wave contribution for both parallel and  
 normal motion the non-local optic effects are
negligibly small for practically all separations. For normal relative 
motion for $p$-wave contribution the non-local optic effects are less important
than for the parallel relative motion. In the present
calculations we use the non-local dielectric approach which take into 
account the non-local optic effects on the length scale of the skin-depth. There 
are also the short range non-local optic effects coming from the non-local 
nature of the screening response near the surface. 
 This gives the  surface non-local
contribution which we investigated in our previous publications \cite
{Persson and Volokitin,Volokitin2}. Comparing our previous calculations with
the present one, we find that for $d>10\mathrm{\AA }$ the volume
contribution from the non-local effects is of the same importance as the
surface contribution. For $d>10\,$\AA\ the main contribution to the friction
coefficient $\gamma $ comes from $s$-polarized waves. In particularly, at $%
d=100$ \AA\ the $s$-wave contribution $\gamma _s\approx
10^{-5}kgs^{-1}m^{-2} $, so that with the surface area $A\approx 10^{-14}$m$%
^{-2}$ ( typical for probe scanning microscopy), the friction coefficient is
$\Gamma \approx \gamma _sA\sim 10^{-19}kgs^{-1}$. The $s-$wave contribution
is characterized by weak distance dependence for $d<100$\AA , and $\gamma
\sim d^{-2}$ for $d>100$\AA . For good conductors like copper, even for very
short distances, the main contribution to the friction coefficient comes
from the $s-$ polarized electromagnetic waves. This difference between $p-$
and $s-$ polarized waves results from screening effects: Good conductors are
good reflectors for $p-$ polarized field, which implies that they are
ineffective in the emission and absorption of evanescent $p-$ polarized
waves. However these screening effects are less important for $s-$ polarized
waves.

As pointed out in \cite{Volokitin1,Volokitin2,Volokitin3,Volokitin4},
the $p$ -wave contribution increase and the $s-$wave contribution decrease
when the free electron density decrease. Within the local optic
approximation the force of friction diverges when one go to the limit of
zero conductivity. This situation is different from the radiative heat
transfer, where, even in the case of local optics, a maximum in the heat
transfer occurs for conductivities corresponding to semi-metals.
Fig.3 shows the dependence of coefficient of friction on the electron
density. When the electron density decreases there is transition from a
degenerate electron gas to a non-degenerate electron gas at the density $%
n_F\sim (k_BTm)^{3/2}/\pi ^2\hbar ^3$. At $T=273\,K$ the transition density $%
n_F\sim 10^{25}\,m^{-3}$. For $n>n_F$ we use the (non-local) dielectric
function appropriate for a degenerate electron gas, while for $n<n_F$ we use
an expression corresponding to a non- degenerate electron gas. In the
calculations we used the electron mean free path $l\approx 600\,\mathrm{\AA }
$. At $d=100$\AA\   the maximum value $\gamma _{max}\sim 10^{-4}\,kg\cdot s^{-1}$ is obtained
for $n_{max}\sim 10^{22}\,m^{-3}$, corresponding to the DC conductivity $%
\sigma \sim 1(\Omega \cdot m)^{-1}.$
 
\section{Photon tunneling enhancement of the van der Waals friction}

We rewrite the denominator in Eq. (\ref{3eighteen}) in the form
\[
|1-e^{-2kd}R|^4=[(1-e^{-kd}R_r)^2+e^{-2kd}R_i^2]^2
\]
\begin{equation}
\times [(1+e^{-kd}R_r)^2+e^{-2kd}R_i^2]^2
\label{5one}
\end{equation}
where $R_r$ and $R_i$ are real and imaginary part of $R$, respectively 
($R=R_r+iR_i$). Let us suppose that $|R_r|>>R_i$. In this case resonant 
conditions are determined by equation
\begin{equation}
R_r(\omega_{\pm}(q)=\pm e^{qd}
\label{5two}
\end{equation}   
Close to resonance we can write
\[ 
[(1\pm e^{-qd}R_r)^2\pm e^{-2qd}R_i^2]
\]
\begin{equation}
\approx e^{-2qd} R_r^{\prime 2}(\omega_{\pm})[(\omega-\omega_{\pm})^2+
(R_i(\omega_{\pm})/ R_r^{\prime}(\omega_{\pm}))^2]
\label{5three}
\end{equation}
where
\[
R_r^{\prime}(\omega_{\pm})=\left.\frac{dR_r^{\prime}(\omega)}{d\omega}
\right|_{\omega=\omega_{\pm}}
\]
which leads to the following contribution to the friction coefficient:
\begin{equation}
\gamma_{\pm} \approx \frac{\hbar^2}{16k_BT}\int_0^{q_c}dk\,k^3
\frac{e^{2kd}}{[|R_r^{\prime}(\omega_{\pm})|
R_i(\omega_{\pm})\sinh^2\hbar\omega_{\pm}/2k_BT]}       
\label{5four}
\end{equation}
The parameter $q_c$ in this expression defines the value of $q$ where 
the two poles approximation is valid. To proceed further let us make 
the following simplifications. Close to the poles we can  
 use approximation
\begin{equation}
R=\frac{a}{\omega-\omega_0-i\eta}
\label{5six}
\end{equation}  
where $a$ is a constant. Then from resonant condition (\ref{5two}) we get
\[ 
\omega_{\pm}=\omega_0\pm ae^{-qd}
\]
For the two poles approximation to be valid the difference $\Delta\omega =
|\omega_+-\omega_-|$ must be greater than the width of the resonance $\eta$. 
 From this condition we get
$q_c\le\ln(2a/\eta)/d$.
For the short distances the parameter $q_c$ defines the value of $q$ where 
the solution of Eq. (\ref{5two}) ceases to exit. 

For $\omega_0>a$ and $q_cd>1$ from Eq. (\ref{5four}) we get
\begin{equation}
\gamma_{\perp\pm}=\frac{3}{128}\frac{\hbar^2a^2}{d^4k_{B}T\eta}
\frac{1}{\sinh^2(\hbar\omega_0/2k_BT)}
\label{5seven}
\end{equation}

For the parallel relative motion using the same approximation we get
\begin{equation}
\gamma_{\parallel}=\frac{\hbar^2 \eta q_c^4}{128\pi k_{B}T}
\frac{1}{\sinh^2(\hbar\omega_0/2k_BT)}
\label{5eight}
\end{equation}    
Interesting, the explicit $d$ dependence has dropped out of Eq. (\ref{5eight}). 
However, $\gamma_{\parallel}$ is still $d$ dependent, through the $d$ dependence 
of $q_c$. For the small distances one can expect that $q_c$ is determined 
by the dielectric properties of the material and does not depend on $d$. In this 
case the friction will be also distance independent. Probably, the weak distance 
dependence observed in \cite{Stipe} can be explained by the resonant photon 
tunneling.

Resonant photon tunneling enhancement of the van der
Waals friction is possible for two semiconductor surfaces which can support
low-frequency surface plasmon modes. As an example we consider
two clean surfaces of silicon carbide (SiC). The
optical properties of this material can be described using an oscillator
model \cite{Palik}
\begin{equation}
\epsilon(\omega)=\epsilon_{\infty}\left(1+\frac{\omega_L^2 - \omega_T^2}
{\omega_T^2 - \omega^2 -i\Gamma \omega}\right)
\label{5nine}
\end{equation}
with $\epsilon_{\infty}=6.7, \,\omega_L=1.8\cdot10^{14}s^{-1},\,
\omega_T=1.49\cdot10^{14}s^{-1},\,$ and $\Gamma=8.9\cdot10^{11}s^{-1}.$ The
frequency of surface plasmons is determined by condition
$\epsilon_r(\omega_p)=
-1$ and from (\ref{three}) we get $\omega_p=1.78\cdot10^{14}s^{-1}$. In Fig.2
we plot the friction coefficient $\gamma(d)$: note that the
friction between the two semiconductor surfaces is  several order of magnitude
larger than between two clean  good conductor surfaces.

Another enhancement mechanism is connected with resonant photon tunneling
between adsorbate vibrational modes localized on different surfaces. 
As an example, let us consider ions with charge $e^*$ adsorbed on metal surfaces.
The polarizability for ion vibration normal to the surface is given by
\begin{equation}
\alpha_{\perp}=\frac {e^{*2}}{M(\omega_{\perp}^2-\omega ^2 -i\omega \eta_{\perp})},
\end{equation}
where $\omega_{\perp}$ is the frequency of the normal adsorbate  vibration,
and $\eta_{\perp}$ is the damping constant.
In Eq. (\ref{3twentyone}) the contribution  from parallel vibrations
is reduced by the small factor $1/\epsilon$. However, the contribution
of parallel vibrations to the van der Waals friction
 can nevertheless   be important
due to the indirect interaction of parallel adsorbate vibration with the electric field,
via the metal conduction electron \cite{Persson and Volokitin2}. Thus, the small
parallel component
of the electric field will induce a strong electric current in the metal.
The drag force between
the electron flow and adsorbates can induce adsorbate vibrations  parallel to the
surface. This gives the polarizability:
\begin{equation}
\alpha_{\parallel}=\frac {\epsilon -1}{n}\frac {e^{*}}{e}\frac {\omega \eta_{\parallel}}
{(\omega^2_{\parallel}-\omega^2 -i\omega \eta_{\parallel})}
\end{equation}
where $n$ is the conduction electron concentration.
As an illustration, in Fig.3 we show coefficient of friction for the two
Cu(001) surfaces covered by a low concentration of potassium atoms
( $n_a=10^{18}m^{-2}$)
. In  the $q-$
integral in Eqs.(\ref{3eighteen},\ref{3twenty}) we used the cut off  
$q_c \sim \pi/a$ (where $a\approx1nm$
is the inter-adsorbate
distance) because our microscopic approach is applicable only when the wave length
of the electromagnetic field is larger than double average distance between the
adsorbates.
 In comparison, the friction
between two clean surface at separation $d=1nm$ is seven order of
magnitude smaller.
 At $d=1nm$
the friction coefficient $\Gamma$ for an atomic force microscope tip
with $R \sim 1\mu m$
is $\sim 10^{-12}kgs^{-1}$ ($\gamma \sim 10^{3}kgs^{-1}m^{-2}$, see
Fig.2);
this is of the same order of magnitude
as the  observed friction \cite{Stipe}.

\section{Summary and conclusion}

We have calculated the van der Waals friction between two flat surfaces 
for normal relative motion and found a drastic difference in the comparison 
with parallel relative motion. This difference is connected with resonance 
produced by the multiple scattering of the electromagnetic waves from the opposite 
surfaces. In the case of sharp resonance it gives much larger contribution to 
friction in the case of normal relative motion than for parallel relative 
motion. 

We have studied in the detail the friction between two good conductors and have 
found that for normal relative motion even for very small distances the main 
contribution to friction comes from the retardation effects. The non-local optic 
effects are very important for $p$-wave contribution to friction for parallel 
relative and much less important for normal relative motion. For $s$-wave 
contribution the non-local optic effects are unimportant for both direction 
of relative motion.

In the case of van der Waals friction we have found that for distances
between bodies $\sim 100$\AA ,\ for good conductors with a high free
electrons concentration, the main contribution to friction is associated
with the $s$-polarized electromagnetic waves. For $d<100$\AA\ this mechanism
gives a friction coefficient per unit area $\gamma \sim 10^{-5}$kgs$^{-1}$m$
^{-2},$ nearly independent of the distance $d,$ while for $d>100$\AA\ the
friction coefficient $\gamma $ depends on distance as $d^{-2}.$ For an
atomic force microscope tip with the near substrate area $A\sim 10^{-14}$m$
^2,$ we got  the friction coefficient $\Gamma \sim \gamma A\sim 10^{-19}$kgs
$^{-1}$ for $d<100$\AA. When the concentration of electrons decreases,
the $s-$contribution to the friction decreases while the $p-$ contribution
increases. At $d=100$\AA\   and with the electron lifetime $\tau = 4 \cdot
 10^{-14}$s, the $p-$ contribution reaches maximum $\gamma _{\max }\sim
10^{-4}$kgs$^{-1}$m$^{-2}$ at the electron concentration $n\sim 10^{22}$m$
^{-3},$ which corresponds to the conductivity $\sigma \sim 1\,$($\Omega $m)$
^{-1}.$

We have shown that the van der Waals friction can be enhanced by several orders of 
magnitude in the case of resonant photon tunneling between low-frequency 
surface plasmon modes and adsorbate vibrational modes. In the case of friction 
for two Cu(100) surfaces covered by a low concentration of potassium atoms 
at $d=10\AA$ we have found friction of the same order of magnitude as it was 
observed in experiment \cite{Stipe}. However, the distance dependence in this 
case is more stronger than it was observed in \cite{Stipe}. Further experiments 
with simple and well defined composition of the tip and sample must be performed 
to elucidate different energy dissipation mechanisms in the non-contact friction.

The obtained results should have broad application in non-contact friction microscopy, 
and in design of new tools for studying adsorbate vibrational dynamics and optical 
properties of the surface plasmons.       
  
\vskip 0.5cm \textbf{Acknowledgment }

A.I.V acknowledges financial support from DFG and the Russian Foundation for
Basic Research (Project No. 01-02-16202) B.N.J. P acknowledges financial
support from BMBF. We thank R.O. Jones for help in the numerical
calculations.\vskip 1cm

\appendix

\section{\,\,\,}
 After substituting
(\ref{one}) and (\ref{two}) into formula (\ref{3thirteen}) to linear order in
vibrational coordinate $u_0$ and frequency $\omega_0$ we get
\[
\sigma _{zz}=\frac 1{4\pi }\int_0^\infty d\omega \int \frac{d^2q}{(2\pi)^2}
\Big(\frac{p}{q^2}\Big[(p+p^{*})(<\left|w_{0z}\right|^2 + <\left|v_{0z}\right|^2>
)
\]
\[
 + (p-p^{*})(<w_{0z}v^{*}_{0z}> + c.c)] +
\left(\frac{c}{\omega}\right)^2p\Big[(p+p^{*})(<\left|w_{0y}\right|^2>
 + <\left|v_{0y}\right|^2>)
\]
\[
+(p-p^{*})(<w_{0y}v^{*}_{0y}> + c.c)\Big]
+\Big(\frac{p^+}{q^2}\Big[(p+p^{*})(<w_{1z}w_{0z}^{*}>+<v_{1z}v_{0z}^{*}>+c.c.)
\]
\[
  (p-p^{*})(<w_{1z}v_{0z}^{*}>+<v_{1z}w_{0z}^{*}>+c.c.)\Big]+\frac{c^2}{\omega(\omega
+\omega_0)}p^{+}\Big[(p+p^{*})(<w_{1y}w_{0y}^{*}>
\]
\begin{equation}
+<v_{1z}v_{0z}^{*}>+c.c.)+(p-p^{*})(<w_{1y}v_{0y}^{*}>+<v_{1y}w_{0y}^{*}>+c.c.)\Big]
\Big)e^{-i\omega_0t}\Big)
\label{aone}
\end{equation}

From Eqs. (\ref{3fifteen}) and (\ref{aone}) it follow that the friction coefficient is determined by the formula
\[
\gamma_{\perp}=\frac 1{4\pi u_0i }\int_0^\infty d\omega_0 \int \frac{d^2q}{(2\pi)^2}
\Big(\frac{\partial}{\partial \omega_0}\Big(\frac{p^+}{q^2}\Big[(p+p^{*})
(<w_{1z}w_{0z}^{*}>+<v_{1z}v_{0z}^{*}>-c.c.)
\]
\[
  (p-p^{*})(<w_{1z}v_{0z}^{*}>+<v_{1z}w_{0z}^{*}>-c.c.)\Big]+\frac{c^2}{\omega(\omega
+\omega_0)}p^{+}\Big[(p+p^{*})(<w_{1y}w_{0y}^{*}>
\]
\begin{equation}
+<v_{1z}v_{0z}^{*}>-c.c.)+(p-p^{*})(<w_{1y}v_{0y}^{*}>+<v_{1y}w_{0y}^{*}>-c.c.)\Big]
\Big)\Big)_{\omega_0=0}
\label{atwo}
\end{equation}
Using Eqs. (\ref{nine}-\ref{thirteen},\ref{corfs},\ref{corfp}) we get
\[
\frac {1}{q^2}\frac{\partial}{\partial \omega_0}\left(p^+(p+p^*)(<w_{1z}w_{0z}^*>+
<v_{1z}v_{0z}^*>)-c.c.\right)_{\omega_0=0}=2i u_0\left(n(\omega)+\frac{1}{2}\right)   
\]
\begin{equation}
\times \frac{\partial}{\partial 
\omega}p^2\Big[\frac{(1-|R_{1p}R_{2p}|^2)^2+
|(1-|R_{1p}|^2)R_{2p}e^{ipd}+(1-|R_{2p}|^2)R_{1p}^*e^{-ipd}|^2}
{|\Delta_p |^4}\Big]   \label{athree}
\end{equation}
\[
\frac{1}{q^2}\frac{\partial}{\partial \omega_0}\left(p^+(p-p^*)(<w_{1z}v_{0z}^*>+
<v_{1z}w_{0z}^*>)-c.c.\right)_{\omega_0=0}=8i u_0\left(n(\omega)+\frac1{2}\right)   
\]
\[
\times \frac{\partial}{\partial 
\omega}\frac{p^2}{|\Delta_p |^4}\Big[(\mathrm{Im}R_{1p} 
+e^{-2|p|d}|R_{1p}|^2\mathrm{Im}R_{2p})
(\mathrm{Im}R_{2p}+e^{-2|p|d}|R_{1p}|^2\mathrm{Im}R_{2p})   
\]
\begin{equation} 
+e^{-2|p|d}\mathrm{Im}(R_{1p}R_{2p})^2
\Big]e^{-2|p|d}   \label{afour}
\end{equation}
Other similar expressions for the $s$-wave contribution  can be obtained from 
Eqs. (\ref{athree}) and (\ref{afour}) by replacement of the reflection factors
$R_p$ for $p$-polarized wave by the reflection factors $R_s$ for $s$-polarized 
wave. After substituting Eqs. (\ref{athree}) and (\ref{afour}), and similar
expression for $s$-polarized waves in Eq. (\ref{atwo}) we get the friction coefficient
for normal relative motion which is determined by formulas (\ref{3seventeen}-
\ref{3eighteen}).

\thinspace \thinspace \thinspace

\vskip 0.5cm

FIGURE CAPTIONS

Fig. 1. The friction coefficient for two flat surfaces in parallel relative 
motion as a function of
separation $d$ at $T=273\,$K with parameter chosen to correspond to copper ($%
\tau ^{-1}=2.5\cdot 10^{13}s^{-1},\,\omega _p=1.6\cdot 10^{16}s^{-1}$). The
contributions from the $s-$ and $p-$polarized electromagnetic field are
shown separately. The full curves represent the results obtained within the
non-local optic dielectric formalism, and the dashed curves represent the
result obtained within local optic approximation. (The log-function is with
basis 10)

Fig. 2. The friction coefficient for two flat surfaces in normal relative 
motion as a function of
separation $d$ at $T=273\,$K with parameter chosen to correspond to copper ($%
\tau ^{-1}=2.5\cdot 10^{13}s^{-1},\,\omega _p=1.6\cdot 10^{16}s^{-1}$). The
contributions from the $s-$ and $p-$polarized electromagnetic field are
shown separately. The full curves represent the results obtained within the
non-local optic dielectric formalism, and the dashed curves represent the
result obtained within local optic approximation. (The log-function is with
basis 10)

Fig. 3. The friction coefficient for two flat surface as a function of the
free electron density $n$ at $T=273\,$K. The full curve was obtained by
interpolation between the result (dashed lines) obtained within the
non-local optic dielectric approach, with dielectric functions corresponding
to a degenerate electron gas for $n>n_F\sim 10^{25}$m$^{-3}$ , and to a
non-degenerate electron gas for $n<n_F.$ The calculation were performed with
the damping constant $\tau ^{-1}=2.5\cdot 10^{13}s^{-1}$ , separation $d=100$%
\AA\ and $n_0=8.6\cdot 10^{28}m^{-3}.$ (The log-function is with basis 10)

Fig.4. The friction coefficient for two  clean semiconductor surfaces
in (a) normal and (b) parallel relative motion, as a function of the
separation $d$. $T=300$K and  with parameters chosen to correspond to
a surfaces of silicon carbide (SiC) (see text for explanation)
(The log-function is with basis 10)

Fig. 5. The friction coefficient for two  surface covered by adsorbates
 in (a) normal and (b) parallel relative motion, as a function of the
separation $d$. $T=273$K and  with parameters chosen to correspond to K/Cu(001)
\cite{Senet} ($ \omega_{\perp}=1.9\cdot 10^{13}s^{-1}, \omega_{\parallel}=
4.5\cdot 10^{12}s^{-1}, \eta_{\parallel}=2.8\cdot 10^{10}s^{-1},
 \eta_{\perp}=1.6\cdot 10^{12}s^{-1}, e^{*}=0.88e$)
(The log-function is with basis 10)

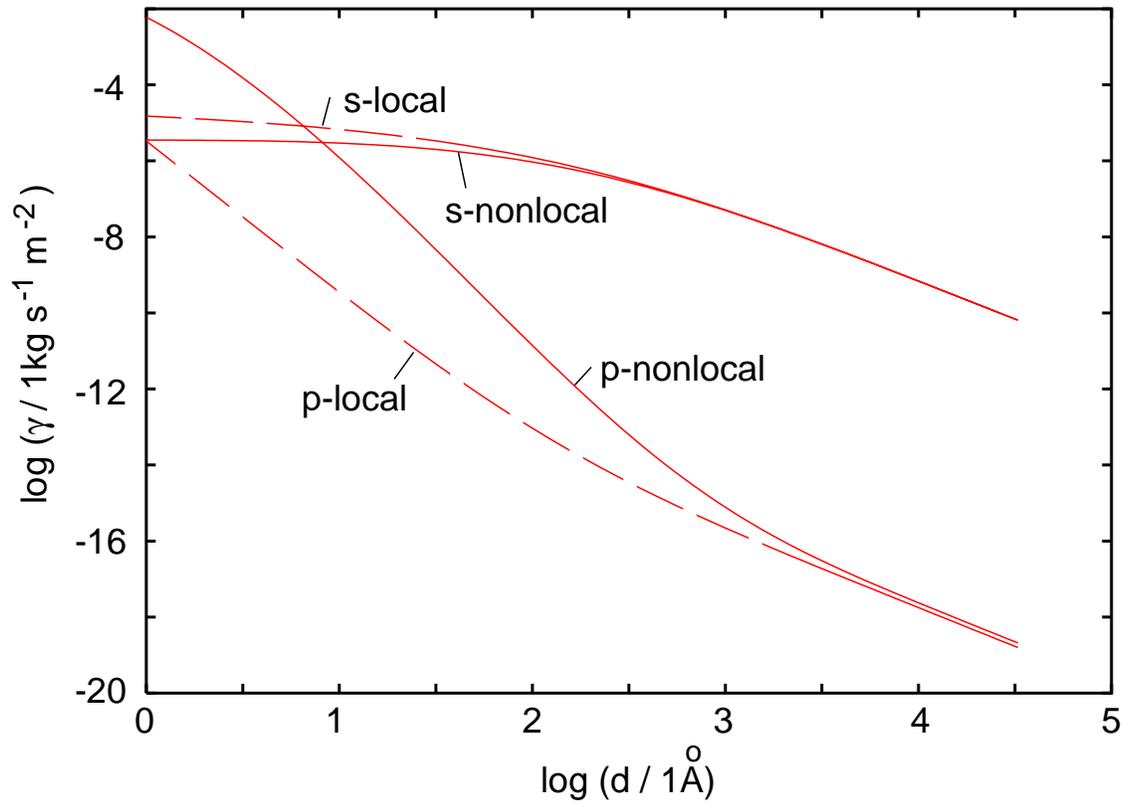
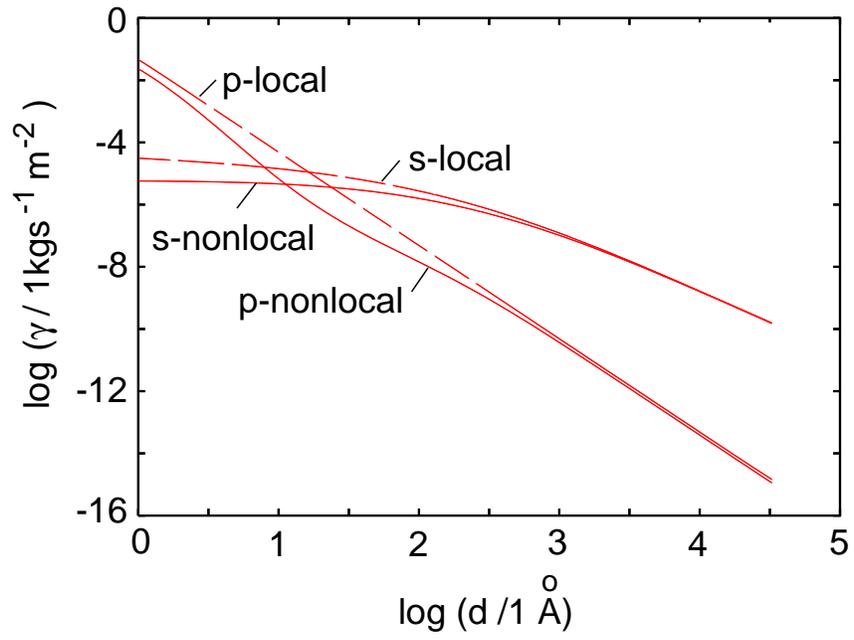
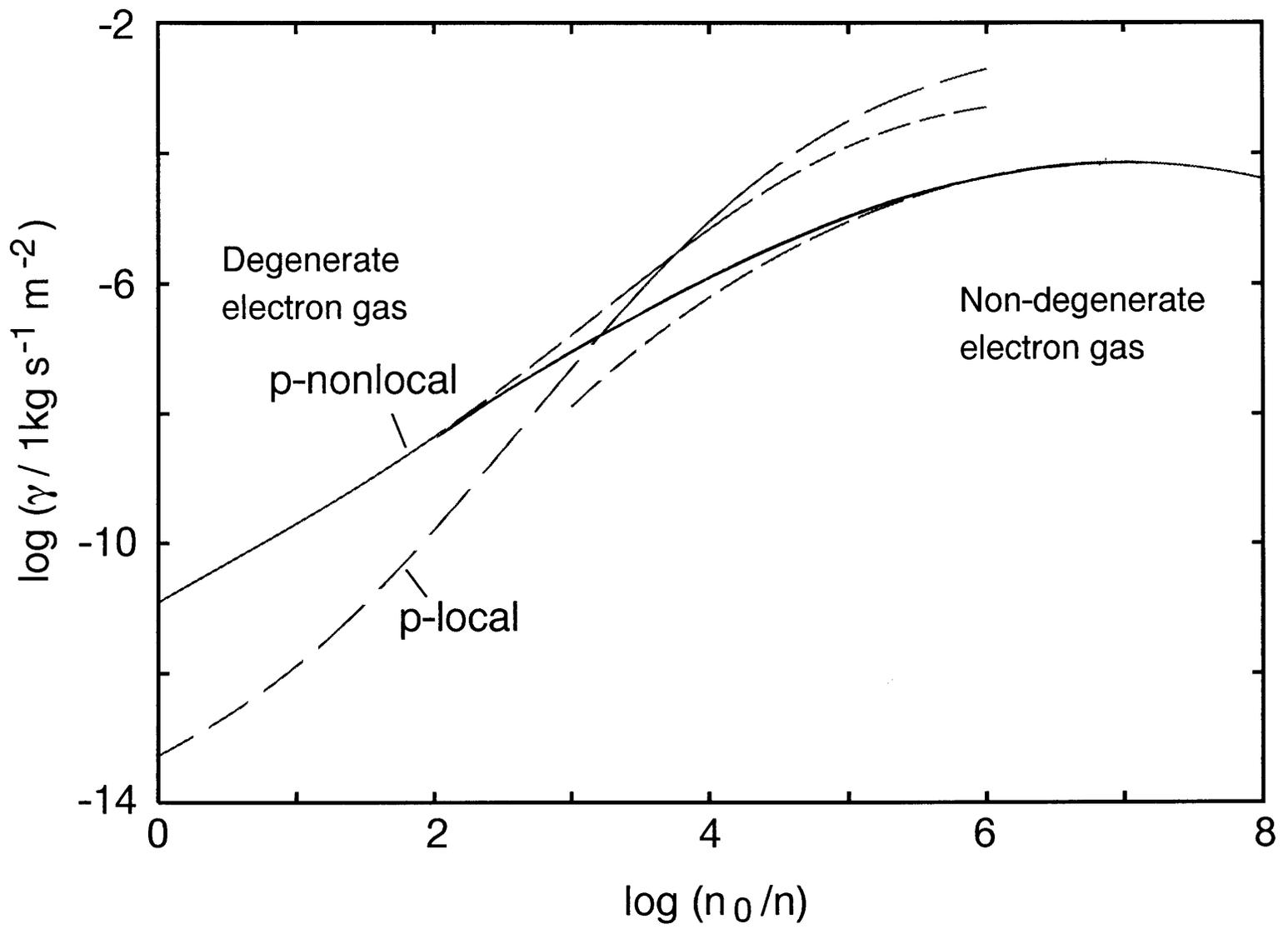
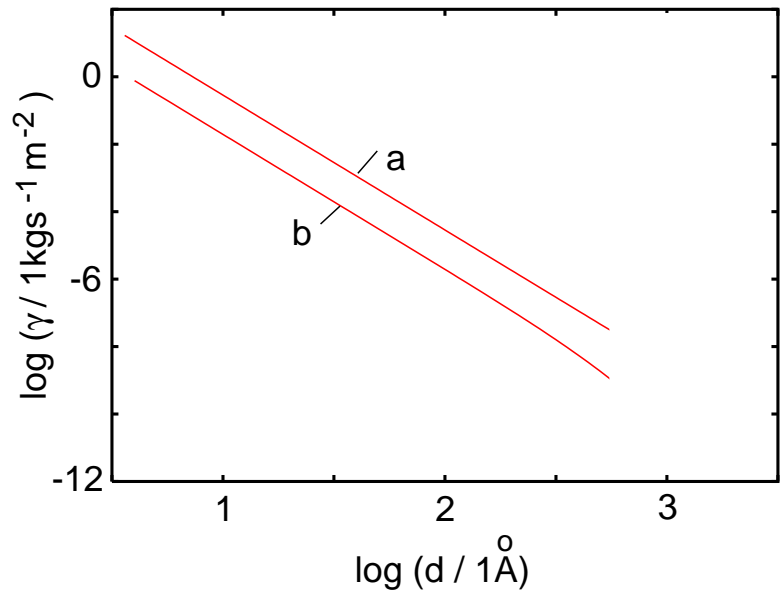
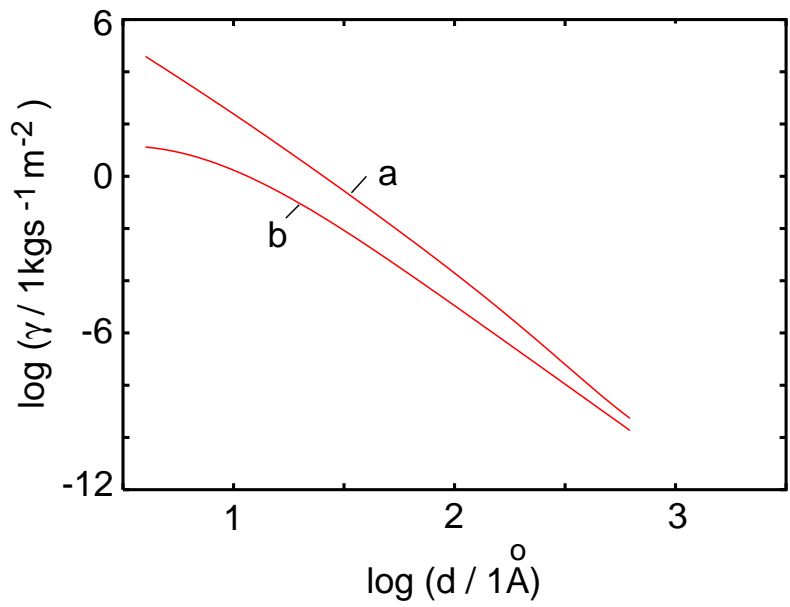
\end{document}